\RequirePackage[hyphens]{url}
\documentclass[pageno]{jpaper}

\usepackage{float}
\usepackage{ulem}
\usepackage{booktabs}
\usepackage{mathtools}
\usepackage{ragged2e}
\usepackage{listings}
\usepackage{balance}
\usepackage{soul}
\usepackage{etoolbox}
\usepackage{mdframed}
\usepackage{xcolor}
\usepackage{varwidth}
\usepackage{authblk}
\newfloat{code}{H}{prg}
\floatname{code}{Code Listing}
\newcommand{\todo}[1]{}
\renewcommand{\todo}[1]{{\color{red} ({#1})}}

\begin{document}
\title{Open-source Hardware: Opportunities and Challenges}

\author[1,2]{
Gagan Gupta
  }
\author[2]{
Tony Nowatzki
  }
\author[2]{
Vinay Gangadhar
  }
\author[2]{
Karthikeyan Sankaralingam
  }
\date{\centerline{\{gagang, tjn, vinay, karu\}@cs.wisc.edu}}
{ 
  \affil[1]{Microsoft Research}
  \affil[2]{
  University of Wisconsin - Madison 
  }

\maketitle
\begin{abstract}
Innovation in hardware is slowing due to
rising costs of chip design and diminishing benefits from Moore's law
and Dennard scaling. Software
innovation, on the other hand, is flourishing, helped in good measure by
a thriving open-source ecosystem. We believe that open
source can similarly help hardware innovation, but has not
yet due to several reasons.  We identify these reasons and how the
industry, academia, and the hardware community at large can come
together to address them.  
\end{abstract}

\section{Introduction} \label{sec:intro}

Advances in silicon technology and hardware architecture have been crucial in
enabling new computing technologies and applications.
However, current trends, the slowing of Moore's law and Dennard scaling, are
slowing silicon technology advances. 
Meanwhile, hardware design has become ever more complex and expensive, especially chip design.
Consequently, innovation in hardware is slowing.  By several accounts, the semiconductor
industry's growth is slowing and the number of new hardware
startups is dwindling~\cite{noonen}.
Unfortunately this is happening at a time when innovation is needed in many areas, e.g., big data processing, machine learning, augmented reality, etc.

In contrast, innovation and revenue-growth in software is flourishing.
According to Software \& Information Industry
Association report, private firms reported annual average growth of
55\% for 2014 (\url{https://www.siia.net/Press/Software-Industry-Revenue-Growth-Accelerating-and-Hiring-/Expected-to-Jump-According-to-New-SIIA-OPEXEngine-Report}).
Google, Facebook, and Twitter are but a few mega-success stories of the last decade.
Uber, Pinterest, and Airbnb are examples of some new successful startups.
In 2013, software startups attracted fifteen-fold more investment than hardware startups~\cite{noonen}.

We observe that software-based ventures aggressively leverage the thriving \emph{open-source software} (OSS) ecosystem to build products and services.
For example, Facebook started with PHP, Twitter and Shopify use
Ruby on Rails, Uber uses Node.js, and Pinterest uses Hadoop and
Memcached, all open-source platforms.  Using open-source technology helps
innovate faster, shorten time to market, and minimize investment~\cite{startup_cost}.
This improves chances of 
success, fueling more innovation in turn. 

\begin{figure}[t]
\begin{center}
    \includegraphics[width=\linewidth]{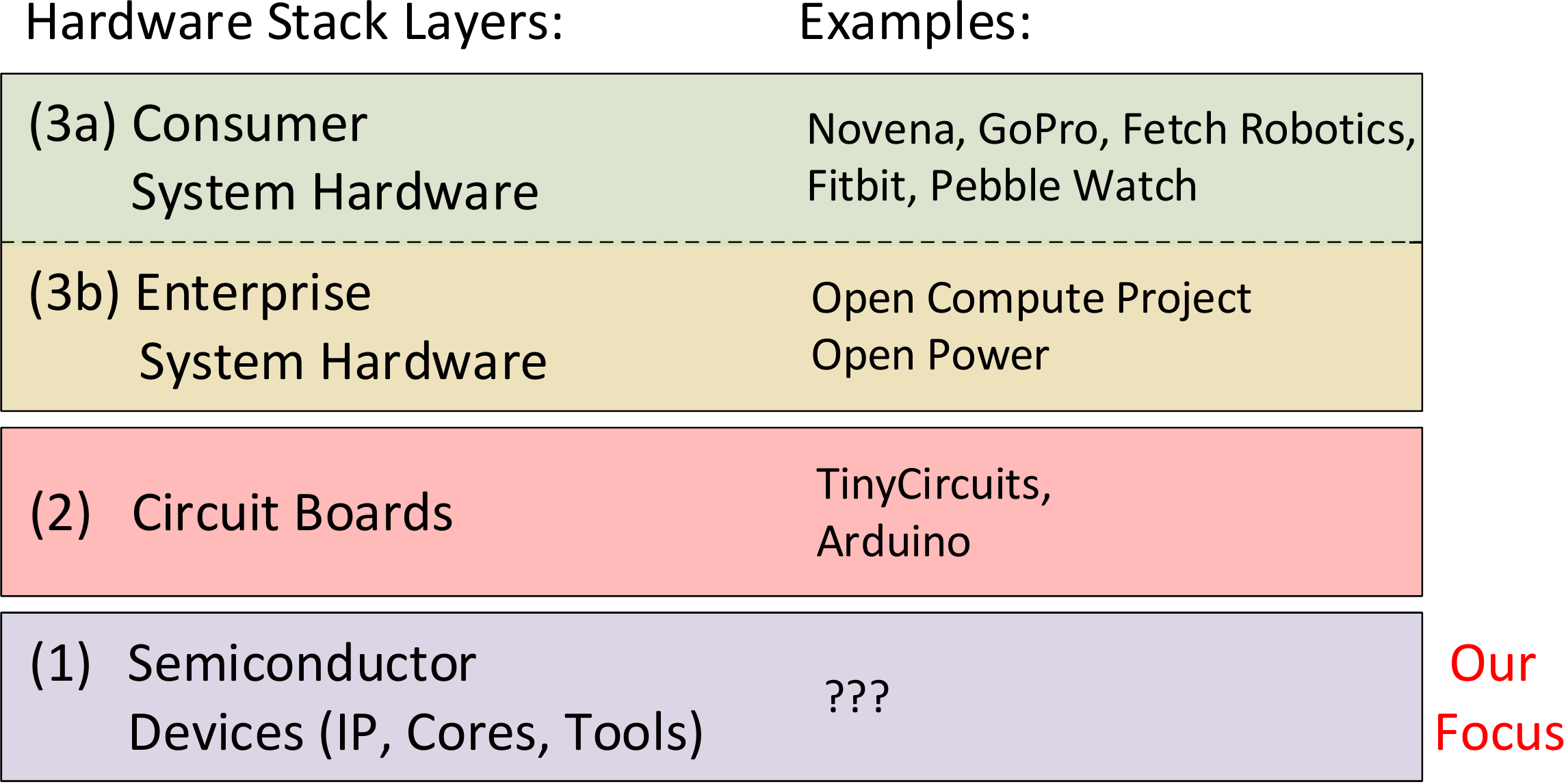}
    \vspace{0.03in}
    \caption{Open-source Success Stories in Hardware}
    \vspace{-0.11in}
    \label{fig:ecosystem}
\end{center}
\end{figure}

Open source has not pervaded the hardware industry in
an analogous way.
While open source has been fruitful at the
\emph{system} hardware and \emph{circuit board} levels (see
Figure~\ref{fig:ecosystem}), it has been inconsequential 
at the \emph{semiconductor} level for SoC and FPGA design.
This is where innovation is now needed the most, but has stalled.
We believe that open-source hardware (OSH) can drive SoC/FPGA innovation and industry growth, by enabling
wider community participation, low-cost development, and quick adoption of new
device-level technological breakthroughs.

Others have also argued that OSH can enable SoC/FPGA innovation ~\cite{moores_death_spur}, 
but no one has yet identified a practical path forward.
Hence here we address the questions: \emph{Why is OSH not as successful as OSS?}
\emph{How can we kick-start a vibrant OSH movement?}
 
In Section~\ref{sec:cycle} we first examine the benefits of open source to  the software industry, and
the differences and challenges to OSH.
We then identify how academia, industry, and hobbyists can forge a substantive OSH movement, by leveraging emerging technological trends, e.g., new FPGA platforms and stabilizing technology nodes, and emerging social trends, e.g., easier access to industry resources and new academic OSH efforts. 
Section~\ref{sec:case} shows how OSH can enable rapid innovation in a hypothetical product for face recognition.
Some readers
may skip to Section~\ref{sec:case} for insight on how OSH can be useful in practice, before returning to Section~\ref{sec:background}.

\section{Background}\label{sec:background}

We begin with brief background on two pieces that are likely familiar to
many readers: hardware design process and nascent OSH efforts.

\subsection{Semiconductor Hardware Design}
Designing chips is complex and follows three broad steps as outlined in
Figure~\ref{fig:hw-development}: \textbf{front-end design},
\textbf{back-end design}, and \textbf{fabrication}. An array of EDA (Electronic Design Automation)
tools are used at each step.

During \emph{front-end design}, a solution is architected and is implemented, typically in RTL (Register Transfer Logic).
The design may incorporate pre-existing components like memory and bus controllers.
Some components, e.g, SRAMs and I/O pads, may only be behaviorally modeled in RTL.
Once developed, the RTL is extensively tested 
and verified.
Verification often dominates the effort and cost of front-end design.

\emph{Back-end design} transforms the RTL into a physical design.
The RTL is first synthesized into a gate-level netlist.
Physical counterparts of components that were only modeled in RTL, e.g., the SRAMs and I/O pads, are incorporated in the netlist. 
The netlist is made testable.
Then the gates are physically placed and wires connecting them are laid out. 
The physical design must adhere to strict design guidelines to ensure integrity and manufacturability.
Hence, the physical design is also verified at different stages by performing timing, thermal, power, EM/IR (Electromigration and Voltage drop), ESD (Electorstatic Discharge), parasitic, test vector, LVS (Layout vs Schematic), and DRC (Design Rule Checking) analysis.

Complex components, e.g., processor cores and SRAMs, also referred to as IP (intellectual property),
are typically designed by experts and reused.

In the final, \emph{fabrication step}, masks are made from the physical design in the 
form of GDSII (Graphical Database System II) and the design is fabricated on silicon wafers.
The silicon wafer is cut into die pieces and tested. 
Functional die is packaged  and tested again, before being shipped.

Specialized EDA tools are needed in each step.
For example, simulators are needed to verify the RTL, synthesis tools to generate the netlist, and 
place-and-route tools to place and lay out the design.
Physical design verification tools are needed to check the design's integrity.
Fabrication and packaging are highly specialized processes, often performed by experts other than the designers.

\begin{figure}
  \begin{center}
  \includegraphics[width=1.0\linewidth]{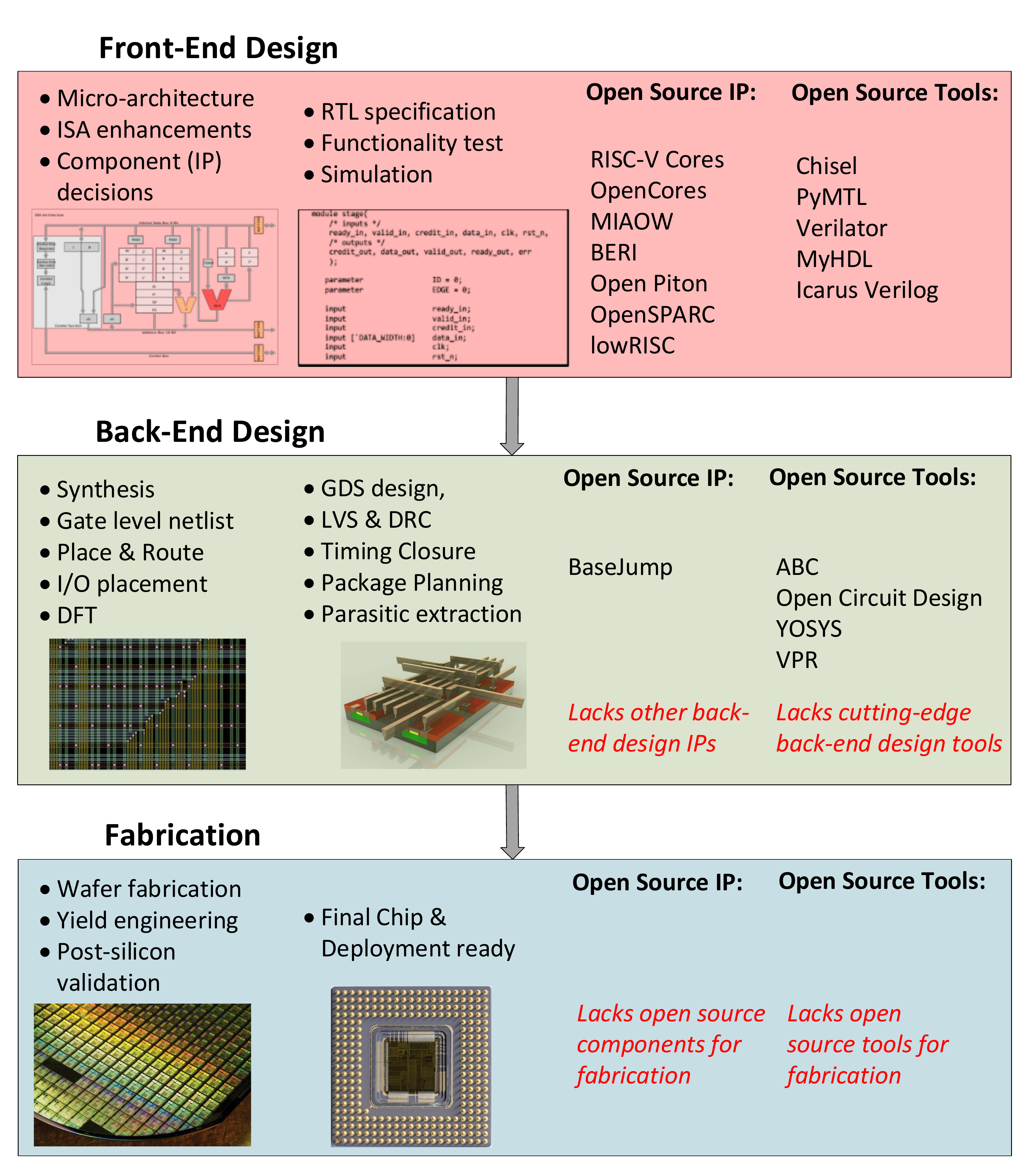}
  \vspace{-0.1in}
  \caption{Semiconductor Hardware Development Steps \textnormal{\small }  }
  \label{fig:hw-development}
  \end{center}
\end{figure}

\subsection{Open-source Hardware Efforts}
While not as widespread as OSS, prevailing OSH
efforts target each step of the design process,
as outlined in the last column of Figure~\ref{fig:hw-development}.
Front-end efforts are plentiful with full-fledged cores like
RISC-V Rocket (~\url{www.riscv.org}), OpenRISC, BERI, Open Piton many-core processor, OpenSPARC, LEON and
GPUs like MIAOW~(\url{www.miaowgpu.org}), Nyami, and Nyuzi~(\url{www.nyuzi.org}). The {\url {opencores.org}} repository
provides a wide assortment of modules, e.g., memory controllers,
ALUs, floating-point units, USB controllers and Ethernet controllers.
lowRISC (~\url{www.lowrisc.org}) is an open-source SoC effort.
Since \emph{back-end} modules like SRAMs and I/Os are
closely tied to a foundry and technology node, building these
requires specialized knowledge that is often proprietary. Hence there
are fewer back-end OSH modules. There are even fewer fabrication- and
packaging-related OSH efforts, although BaseJump (\url{bjump.org}) is 
an example from academia.  

On occasion industry provides free IP for prototyping, e.g., ARM provides Cortex-M0 microcontroller and foundries like TSMC provide back-end IP, but not as modifiable open source.
  
  EDA tools developed by the community include front-end tools like
Verilator, new front-end languages like Chisel and PyMTL, and back-end
tools like Berkeley ABC, Open Circuit Design, YOSYS, and VPR.
Because of the sophistication of
foundry design rules, back-end tools are targeted at older technology
nodes.
OpenAccess is an industry-supported open-source program for EDA tools.

\newenvironment{tightemize}{\vspace{-0.09in} \begin{itemize}}{\vspace{-0.10in} \end{itemize} }

\begin{table*}[tbp]
\setitemize{labelindent=0.02in,labelsep=0.07in,nolistsep} 
\centering
\small
\caption{Differences between hardware and software. Role of industry and community in enabling open-source hardware.}
\label{tbl:summary}
\setlength{\tabcolsep}{4pt}
\begin{tabular}{@{}>{\RaggedRight}p{0.59in}>{\RaggedRight}p{1.3in}>{\RaggedRight}p{1.62in}>{\RaggedRight}p{1.7in}>{\RaggedRight}p{1.35in}@{}}
\toprule
& \multicolumn{1}{c}{Meaningful}
& \multicolumn{1}{c}{Practical}
& \multicolumn{1}{c}{Critical Mass}
& \multicolumn{1}{c}{Deployment} \\ \midrule
HW vs SW 
Differences &
\begin{tightemize}\item Fewer developers and design complexity  \end{tightemize}& 
\begin{tightemize}\item Lack of cheap development infrastructure \end{tightemize}& 
\begin{tightemize}\item Lack of IP and platforms \end{tightemize}&
\begin{tightemize}\item Non-zero deployment cost \end{tightemize} \\ \midrule \midrule
Community Role & 
\begin{tightemize}\item Develop more components and increase usability \end{tightemize} & 

\begin{tightemize}\item Develop effective design tools \item Provide free FPGA farms  \end{tightemize}  &
\begin{tightemize}\item Build resusable h/w platforms \item Build tools to organize emerging critical mass \end{tightemize} & 
\begin{tightemize}\item Develop tools for post-manufacture testing \end{tightemize}\\ \midrule
Industry Role & 
\begin{tightemize}\item Contribute more  \item Develop alternate business models  \end{tightemize}& 
\begin{tightemize}\item Provide freemium tools \item Reduce cost and simplify path to chip prototypes \end{tightemize}& 
\begin{tightemize}\item Take a leap of faith \& work with OSH community \end{tightemize}&
\begin{tightemize}\item Build customizable appliances \end{tightemize} \\ \bottomrule
\end{tabular}
\end{table*}

\begin{figure}[t]
    \includegraphics[width=0.99\linewidth]{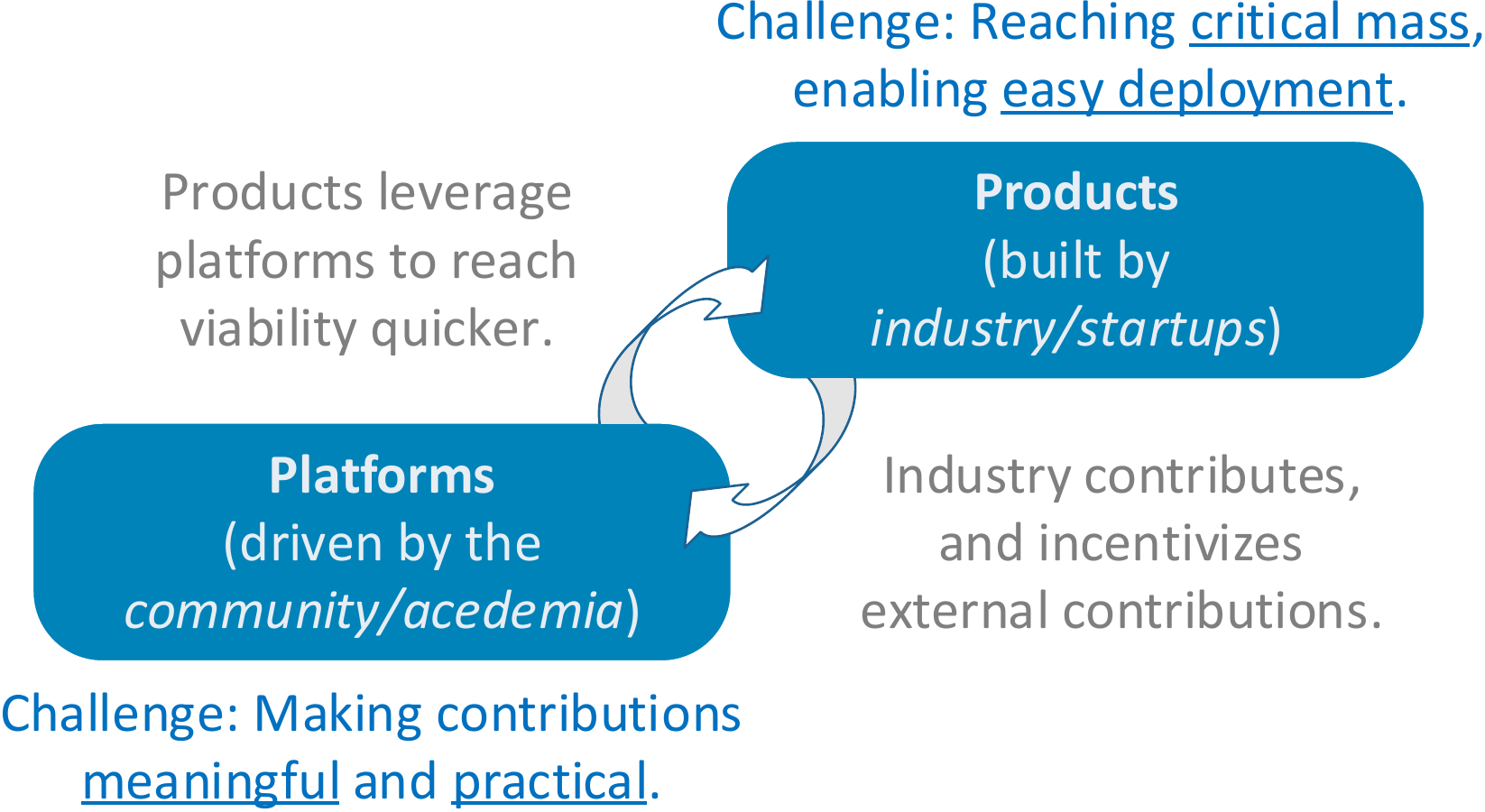}
    \vspace{0.1in}
    \caption{Virtuous Cycle of Platforms to Products}
    \label{fig:virt}
\end{figure}

\section{Stimulating the Open-source Virtuous Cycle}\label{sec:cycle}

Figure~\ref{fig:oss} provides a brief history of OSS
whose evolution and use is instructive to understand the path forward
for OSH. 
In our view, OSS works through a virtuous cycle where a
community of developers build \emph{platforms} with no direct financial
incentive, and industry and startups use these to quickly build
\emph{products}. In turn, industry contributes to the platforms and
incentivises future platform growth (Figure~\ref{fig:virt}).

By analyzing OSS history, we find that an interplay of five ``pillars'' are key to the virtuous cycle of platforms and products.
Developers contribute to open-source platforms when the effort is personally \textbf{meaningful} and \textbf{practical}. 
On the product side, a \textbf{critical mass} of infrastructure significantly lowers the effort to start from an open platform, and \textbf{deploying} software is easy and inexpensive.
Furthermore, permissive \textbf{legal} framework allows developers to contribute and use OSS.

\begin{figure}
\begin{mdframed}[backgroundcolor=black!25]
\begin{minipage}[adjusting]{\linewidth}
\begin{centering}
\caption{Open-source Software, History Primer and Present Use} \label{fig:oss}
\end{centering}
\vspace{0.1in}
\small
Early computer vendors permitted users to modify the software code in their products.
When they stopped this practice, in the mid-80s, Richard Stallman of the MIT AI lab
created open-source software to build systems using entirely free software and
allow users to freely modify the software.
Stallman started with GNU OS, to replace Unix (owned by AT\&T). He
also developed GCC to compile the OS source code, and other utilities (e.g.,
Emacs). In the process others in the AI lab and on usenet groups
contributed.
Serendipitously, Linus Torvalds 
wanted to port MINIX to PC.
Inspired by Stallman, Torvalds used GCC to develop open-source Linux, which became the kernel within GNU OS.
These two developments were critical, if not foundational, in the open-source software movement.

\paragraph{}
Whereas the above developments were academic, 
other open-source software originated because someone wanted to solve a
problem at hand.
For example, Rasmus Lerdorf developed PHP to maintain his personal
homepage, track visitors, and interact with web forms and backend database system,
Yikihiro Matsumoto developed Ruby to create a scripting language better
than Perl and others. David Hansson created Ruby on Rails to develop a custom
web framework. In all these cases, the originators were plugged into a
community on the Internet, they developed fairly extensive first versions,
released the versions for others to use, debug, and contribute. Others found
these tools useful for their own purposes, enhanced them, and contributed the
enhancements back to the pool, perpetuating a cycle in which everyone
benefited.  Once these basic tools were available and attained maturity,
thanks to years of active use and development by a sizable community, 
others like Shopify, Facebook, and Google built successful businesses using the tools. 

\paragraph{}
We cite one example OSS success story, Shopify, an e-commerce company whose recent IPO was 
valued at more than \$1 billion.
Shopify relies heavily on Ruby and Ruby on Rails.
Relying on these platforms helped create a minimum viable product at low cost, estimated at 
\$250,000 to \$300,000~\cite{startup_cost}.
Shopify also perpetuates the OSS cycle by contributing back a number of
Ruby-based projects to make web-development tasks easier (e.g., the
"dashing" app for producing elegant web dashboards).  

\paragraph{}
In considering the history of open-source software, its undeniable that the 
ideological values of freedom, as well as the surrounding culture, were
instrumental in its success.  However, we do not take a position
on the ethical value or moral worth of the \emph{freedom} of software and
hardware for a simple reason:  it's not necessary.  Open source is a driving 
force in the market, which ultimately brings practical value to society.

\end{minipage}
\end{mdframed}
\end{figure}

A similar virtuous cycle has not formed for OSH because OSH differs from OSS, preventing analogous pillars from taking hold. We view the differences and resulting challenges as follows: 
\begin{enumerate}
\item \textbf{Fundamental differences.} These are inherent and arise because
hardware requires physical embodiment (incurring manufacturing cost)
and complex tools (to accomplish multiple non-trivial design processes),
and is inherently concurrent (making it complex to reason about).
\item \textbf{Incidental differences.} Other differences are incidental, arising from 
collective lack of systematic effort, readily-available development platforms, design tools, domain knowledge, and perceived benefits~\footnote{Arguably, fundamental and incidental differences are somewhat interrelated, but we use this categorization to simplify exposition.}.
\end{enumerate}

We analyze the five pillars, their impact on OSS, and the challenges to OSH.
We propose ways in which industry, academics, and hobbyists can together overcome these challenges, including the ones due to the fundamental differences, which perhaps cannot be eliminated, but could be mitigated.
Table~\ref{tbl:summary} provides a summary.

\subsection{Meaningfulness} \label{sec:meaningful}

\paragraph{OSS.} The thriving OSS movement is sustained by individuals
and companies making contributions, because they find it meaningful.
Economists and sociologists have studied this
phenomenon and have identified a few key
motivations~\cite{lakhani2005htu, commercial_oss}. For individuals
they include: (i) intrinsic motivation driven by an ideology of
freedom to share, modify, and distribute software freely; (ii) the
sense of enjoyment or ``flow'' that comes from programming; (iii) skill
development and establishment of credibility from working on OSS
code-bases under scrutiny of peers, which in turn helps in employment
and funding new ventures. The key motivations for industry include: (i)
open-sourcing core technology helps grow business by allowing easy
interoperability with other products; (ii) open-sourcing some
technology helps sell complimentary services for a fee; (iii) it helps
recruit skilled developers familiar with in-house technologies.

\paragraph{OSH.}
Hardware developers, individuals or companies, are not yet as motivated as OSS developers,
due to the fundamental and incidental differences.
Although OSH development does provide intrinsic-motivation and flow, hardware developers are fewer because requiring physical realization can be deterring.
Industry virtually ignores OSH for use in commercial products, and contributes little to front-end,
back-end, or EDA tools, due to lack of perceived value.  The lack of industry recognition
limits OSH participation for skill-development.
Nonetheless, RISC-V conferences and HOTCHIPS 2015, a primarily industry conference, where recent OSH projects on the MIAOW GPGPU and RISC-V processor were presented, show industry's budding interest. 

\emph{Overall, the first challenge is to encourage both the
  development community and hardware industry to more vigorously
  participate in the open-source ecosystem.}

\paragraph{Motivating contributions to OSH\\}
\emph{Community:}
Hardware design may not be immediately usable by the designer, but RTL design is \emph{not} 
inherently difficult to reuse or tweak for different purposes.  
Well-defined interfaces can make components easily interoperable, making the prospect of
sharing meaningful.

Overall, we argue that as components, platforms, and the open-source cycle come to fruition, especially as designing hardware becomes practical (discussed next), the OSH community and its efforts will naturally grow.
The growing maker movement, in which consumers want to build what they use, 
and recent startups, e.g., the ASIC ventures that originated from Bitcoin mining OSH (~\url{https://github.com/fpgaminer/Open-Source-FPGA-Bitcoin-Miner}), will likely motivate more individuals to design chips.
Such efforts can fuel more mainstream consumer hardware devices like the GoPro camera, DJI's drones, and more.

We believe that academia can play a vital role, as they have done with OSS, by structuring courses to attract more students (e.g., by incorporating hands-on hardware projects at college Freshman level~\cite{sigcse2014:arduino} and formulating hardware projects to assist OSH (~\url{http://cseweb.ucsd.edu/~mbtaylor/teaching.html}).

Ultimately, as more and more students, hobbyists, and startups use platforms, there will be more visibility and incentives to contribute.

\emph{Industry:}
Though the OSH movement is likely to grow without the industry's contribution,
their expertise and experience can accelerate the process.
After initial hesitation, 
the software industry found that contributing to OSS was to their
benefit.  
We argue that making commodity, commonly-used  IP freely available
will ultimately benefit the hardware industry.  Proven
IP from vendors, possibly enhanced by the open-source community, and eventually
maintained collectively, can dramatically lower the bar for developers to start
new designs.  
This may disrupt prevailing business models, e.g., IP licensing,
but the industry is likely to respond with alternate business models, like the
software industry did.
Moreover, promoting OSH can benefit ancillary businesses, just as promoting Linux benefited an entire ecosystem.
Ultimately, more design starts will generate more opportunities for the
industry on the whole.

\subsection{Practicality}

\paragraph{OSS.}
The advent of the personal computer and the growth of the Internet in the mid-80s and early 90s made it practical for individuals to contribute to OSS.
As affordable and interconnected PCs proliferated, enthusiasts had the necessary resources to develop software and collaborate.
Moreover, key development tools, such as operating systems like Linux, compilers and debuggers like GCC/gdb, and editors like Emacs, were becoming part of the open source.
These developments enabled the OSS effort and helped it grow.

\paragraph{OSH.} 
Practicality is heavily impacted by hardware's fundamental difference
to software of requiring a physical embodiment and complex (expensive) toolchains.
Neither platforms analogous to the PC, nor tools analogous to GCC are available
to physically realize and design hardware,
posing a significant entry barrier
for hobbyists.

FPGAs mitigate the former concern and alleviate
fabrication availability/difficulty/cost.  However, the FPGA
environment setup is often plagued with tool issues taking weeks and
months, and is rarely like {\tt apt-get module install}. Furthermore,
FPGA designs cannot match chips in area, power, or performance, are on
average 35$\times$ larger than ASICs~\cite{fpgas}, and therefore they may be
suitable for only limited designs.
Although low-cost FPGA platforms, e.g., Hackaday's Arduino-compatible Spartan-6 Shield, are available,
they can house only small designs.
Platforms for larger designs are unaffordable
for students and hobbyists. Regarding complex (expensive) toolchains,
open-source tools like Verilator allow RTL-level simulation,
but the back-end design and fabrication
steps lack cheap and good tools. Licensing commercial
EDA tools and fabricating a prototype can easily cost $\sim$\$1M, well
outside an hobbyist's budget~\cite{adapteva}. Developing back-end
tools and fabrication requires inputs and design rules from foundries, which they seldom disclose.

Further, the domain knowledge and
practice of full chip design is rarely taught formally in
schools and there are only a few in the development community with
requisite experience.

\emph{In short, developing hardware designs today is non-trivial and expensive, 
requiring that the process be simplified and made affordable. }

\paragraph{Making hardware development practical.\\}

We believe that making FPGAs, EDA tools, and fabrication resources easily accessible to developers can make hardware design practical, alleviating challenges inherent to hardware.

\emph{Community:}
A path to making hardware prototyping practical is for FPGAs to become
more ubiquitously available, possibly through general purpose processors (as perhaps Intel's acquisition of Altera portends), 
 mobile platforms (e.g., from Lattice Semi) 
or the cloud (e.g., Microsoft's Catapult).
However, until then, we argue that academia
 host FPGA farms for the general masses.  In academic settings they
can be educational and research tools, and can follow the path of the NEOS
servers (\url{https://neos-server.org/neos/}) which provide state-of-the-art optimization tools at zero cost to
anyone. 
It is also important to simplify the setup and design process, for instance, by
providing device-specific, portable packages that permit users to easily plug
their designs into existing reference designs.

Second, the community, especially academics,
should put a renewed focus on developing efficient and easy-to-use
open-source EDA tools.  In essence, we need the \emph{GCC+glibc+make} of hardware.
There is significant value in creating these
tools, and they would be indispensable going forward.

\emph{Industry:}
As a short-term solution to the lack of good tools, we argue
that the EDA industry make free versions available for non-commercial use, even
if non-premium.  Instead of one-off cases in which designers have to go to
great lengths to acquire free or cheap tools~\cite{adapteva}, we believe that
this should be the norm.  Moreover, as EDA vendors consider the future of the
EDA tools, whether it be to simplify user interfaces, or make the tools
scalable, or deploy them in the cloud, vendors can potentially tap into the
software expertise of the open-source community to evolve their products.
For instance, plugin-enabled tools can attract designers to develop ancillary tools, e.g., timing analysis scripts.
Although this approach runs counter to prevailing practices, more users are likely to create more opportunities for EDA vendors.

Fab shuttle services can lower prototyping costs.
We believe that IP providers
and the handful of EDA vendors and fab houses, can come together
to create a standard operational model that facilitates designers to explore chip
designs.  For example, IP from ARM's DesignStart program, combined with
similar (yet non-existent) standard, (almost) push button EDA flow from Synopsys, for a standard TSMC technology node
(e.g., 45nm), and standard post-manufacturing procedures
can allow developers to go from RTL to prototypes in a
short amount of time with a small investment.

\subsection{Critical Mass}

\paragraph{OSS.} For a commercial venture to use OSS instead of starting from scratch, a critical mass of different platform components has to be available.
For example, today one can build a website by using an open-source operating system (Linux), web framework and language (Ruby on Rails / PHP), memory caching system (Memcached), and backend database (PostgreSQL).
These can usually be combined with very little effort using package-managers. 
This critical mass exists in many areas in software development, simplifying the effort to build minimum viable products (MVPs).

\paragraph{OSH.}
The fundamental differences contribute to OSH's lack of critical mass.
Realizing hardware requires physical components, some of which, like I/O pads and analog IP, are tightly coupled to technology.
They are non-trivial to design, need to be redesigned for each technology node, and hence are not usually openly available.
OSH faces additional issues of not being considered trustworthy and proven, resulting in scant use.

Other incidental issues contribute as well.
Although many components are available from various sources, 
in general, OSH IP has not reached the level of maturity of OSS due to lack of systematic coordination.
However, OSH efforts are gathering momentum. 
For example, the OSHWA (\url{www.oshwa.org}) forum promotes open-source hardware.
RISC-V processor designers have released several cores with accompanying design flows, and more development is in the works. 
Similarly, we have released the RTL and FPGA design flow for MIAOW GPU, and are developing a graphics card, following a standards-based, platform-centric design methodology so that the IP can be easily integrated in other designs.

\emph{Hence the next challenge to OSH is making a critical mass of IP available, that spans from front end to back end.}

\paragraph{Making OSH IP more mature and easily available.\\}

\emph{Community:}
With Moore's law, and hence technology node evolution, slowing~\cite{lastnode}, it
is realistic for the community to develop and maintain back-end IP for the
non-leading edge nodes.

Further, we encourage individuals to contribute towards building platforms, e.g., a functional subsystem comprising a dual-issue processor core, a floating-point unit, a bus controller, a memory controller, and peripherals.
Integrating into a platform helps prove the design's interoperability.
Further, standards-based platforms, e.g, ARM AXI bus-based, are more likely to find users and ensure software portability.
In addition to the front-end design, providing the verification infrastructure and the back-end design flow for the platform will lower the barrier to use it.
Creating dependency managers and packages will facilitate IP sharing/acquisition.

Finally, building trustworthy hardware from untrusted components is an active area of
research, which will help allay the concerns about OSH IP~\cite{Sethumadhavan:2015:THU:2817191.2699412}.

\emph{Industry:}
As a critical mass emerges, at some point industry must take a leap of faith and use OSH, possibly in collaboration with its developers.
Adaptiva's Kickstarter project and Bitcoin miners are encouraging signs of such risk-taking.
We believe that the next step for entrepreneurs is to leverage OSH.
This will also motivate OSH to get to critical mass sooner.

\subsection{Ease of Deployment}
\paragraph{OSS.}
Once developed, software can be simply distributed (over the Internet), 
installed, and upgraded on a host machine with the click of a button.
Developers need develop only for a handful of platforms, mostly based on Windows, Linux, or Mac OS, running on x86 or ARM processors.

\paragraph{OSH.}
Hardware, by nature, cannot match software’s ease of deployment.
Once developed, hardware incurs manufacturing and related costs.
Further, hardware is ``deployed'' through appliances.
As such, there is no analog of  a standard platform on which hardware gets deployed.
However, making deployment easier will encourage
the community to develop a critical OSH mass.

\emph{Hence the next challenge is to reduce the manufacturing cost, and simplify deployment.}

\paragraph{Reducing Manufacturing and Deployment Cost. \\}
Although inherently challenging, several emerging trends can help simplify hardware deployment.

\emph{Community:} The main role of the community in deployment could
be developing tools for post-manufacture test. Secondly, for products
where the deployment is on an FPGA (like Microsoft's Catapult
environment), deployment frameworks analogous to Rails, Heroku and
package managers like Yocto can create a services-layer for hardware.
This perhaps presents new business opportunities for EDA.

\emph{Industry:} OEMs are considering modular appliances in which
hardware components can be swapped like lego pieces~\cite{ara}.  Such
appliances, which also lend well to the maker movement and new
technologies like 3D printing, can simplify hardware deployment.  
We also expect fab shuttle services to become more accessible, e.g., through service providers like eSilicon and MOSIS, as they seek more design starts, and further reduce chip-development cost.

\subsection{Legal Issues}
\paragraph{OSS.}

Open-source licenses play an important role in facilitating the open-source cycle.
In its most basic form, licenses provide a permanent attribution, motivating contributors.
Some restrictive licenses (e.g., GPL) 
require users to share modifications to the OSS as well as their proprietary work that uses the OSS.
This type of license was instrumental in many successful projects, like Linux and GCC, both
directly because industry had to contribute their innovations, and indirectly
because contributors knew that their efforts would not solely benefit corporations.
Less restrictive licenses like LGPL and BSD allow users of OSS to 
not reveal their proprietary components, while the Apache license
expressly permits the use of the licensor's patent rights.
These have proved beneficial for commercial ventures.

\paragraph{OSH.} Since hardware must undergo physical design and
several manufacturing stages, the legal nature of OSH licenses
are more intricate and challenging.  Though OSH contributors today use licenses
such as GPL, BSD, and those from Creative Commons, clear licenses that account
for the hardware design intricacies are lacking~\cite{oshlicense}.

Consider a design that uses proprietary and OSH IPs. 
Just synthesizing the IPs together could be considered a derived
work, possibly exposing the entire design to open-source
licensing, counter to the licensee's intent.
Another example concern is when a developer hands off a design with OSH
components to a contractor, say only for testing.
It is unclear whether this transfer is considered a ``private copy'' or
``distribution,'' and whether it triggers any open-source clauses.

Furthermore, executing complex industry agreements, e.g., non-disclosure agreements (NDAs), can be daunting for academics and hobbyists.

\emph{Therefore, the next challenge is to simplify the legal processes of using OSH.}

\paragraph{Simplifying Legal Processes.\\}
\emph{Community and Industry:} Recognizing the challenges, OSH
promoters like OSHWA (\url{www.oshwa.org}) are formulating licensing
frameworks under which OSH can be disseminated and used.  We believe
that open-source licenses, cognizant of patents, suitable for
 different use models and commercial use need to evolve.
Simple, template agreements from the industry are needed to facilitate prototyping.

\section{The Open-source Cycle in Practice} \label{sec:case}

We have thus far described the open-source cycle in the
abstract.
To make the discussion concrete, we consider one
hypothetical product built using OSH. 

Consider the development of a face-recognition technology for eye-wear, targeting the consumer market like GoPro.
To support real-time operation, new hardware innovation in computational accelerators is necessary.
The design must incorporate an image processing chip for face-recognition, say based on convolution neural networks (CNNs),
alongside other components like a processor core, an image sensor network, memory and bus
controllers, I/O interfaces, etc.
Designing the chip from scratch can take several designers multiple years and cost
millions of dollars, even for a prototype.
Alternatively, we envision that a healthy OSH ecosystem can help develop a MVP
in a much shorter duration and at a lower cost.  

The CNN itself could be first implemented, tested and prototyped using FPGA farms,
open-source tools and OSH components.
Once satisfied with the CNN design, it can be integrated into an OSH platform, e.g., lowRISC's SoC, comprising
a processor core and other components.
The platform can be suitably augmented with additional functionality, e.g., an image sensor network.
Appropriate licenses will enable the platform's use without needing to reveal the CNN design.
Once the design meets its functional specification, free commercial and/or open-source EDA tools
could be used to develop a test-chip, and fabricated using a fab shuttle.
At this point, non-leading edge tools and technology node may suffice.
Platform-related open-source infrastructure, e.g., front-end testbenches and back-end design flows, will considerably simplify the design integration.
Ideally, the only cost incurred would be to fabricate the
test-chip. 
To pursue commercialization,
this ``non-leading edge'' prototype may be productized based on a service model for any of the industry-provided IP and tools. 
Alternatively, the design may be productized using paid-for leading edge IP, tools, and technology node.

\section{Conclusion} \label{sec:conc}

Open-source hardware can accelerate hardware innovation and help
designers build minimum viable products in shorter time and at reduced
costs than is possible today.
We identified five challenges in creating a healthy open-source hardware
ecosystem and outlined a path forward for each.  
First, we
encourage individuals to contribute towards platforms, and provide
the necessary design flows.  We also encourage the industry to
contribute, and contemplate new business models that give free access to commodity IP.
Second, making hardware design practical will encourage more individuals to contribute.
Easily available EDA tools and development platforms, package managers that
simplify IP acquisition and use, and cheap and well-defined
fabrication resources will make hardware design practical.  Third, as more and
more OSH components are created and platforms evolve, a critical mass is likely to
emerge that industry can leverage.  
Fourth, deploying chips in appliances can become easier if OEMs
introduce modular appliances that encourage customization.  Finally,
simple agreements and licenses that grant different degrees of freedom need to evolve, which
will permit commercial and non-commercial parties to use and contribute to
the ecosystem.

A virtuous cycle of open-source hardware platforms and products can
enable innovation and sustainable growth
in the hardware industry.  Academia, industry, and the community at
large have equal roles to play in this endeavor.

\bstctlcite{bstctl:etal, bstctl:nodash, bstctl:simpurl}
\bibliographystyle{IEEEtranS}
\bibliography{osh}
\balance

\end{document}